\documentclass[preprint,showpacs,preprintnumbers,amsmath,amssymb]{revtex4-2}

\usepackage{graphicx}
\usepackage{dcolumn}
\usepackage{bm}

\begin{document}
\title{Spill ripple mitigation by bunched beam extraction with high
frequency synchrotron motion}

\author{S. Sorge} 
\email{S.Sorge@gsi.de}
\author{P. Forck} 
\author{R. Singh} 

\affiliation{%
GSI Darmstadt, Planckstra\ss{}e 1, 64291 Darmstadt, Germany\\
}%

\date{March 7, 2022} 




\begin{abstract}
Slow extraction of bunched beams is widely used for mitigating the formation 
of micro structures due to ripples of the accelerator magnets power supplies 
on the extracted beam which is also referred to as spill. 
That helps, in particular, experiments with slow detectors or with low extraction 
rates. 
Furthermore, this technique is widely used for avoiding long gaps in the spill 
measured in ionisation chambers which in turn would trigger interlocks. 

On the other hand, the bunches create spill structures on time scales defined by the 
rf frequency. 
In addition, macroscopic spill structures of duration up to some hundred 
milliseconds can be created by the synchrotron motion for a sufficiently large synchrotron tune. 

The aim of this report is to study the influence of the synchrotron tune determined by 
amplitude and harmonic number of the rf voltage which additionally depends on 
the transverse emittance and the longitudinal bunch area of the circulating 
beam as well as the distribution of the particles' transit times. 
\end{abstract}

\maketitle

\section{Introduction}

Slow extraction from synchrotrons and storage rings is 
widely used for delivering hadron beams to many fixed target experiments 
as well as for the irradiation of tumours in hadron cancer therapy. 
In most cases, a third integer lattice resonance is excited with 
sextupoles which results in the formation of a triangular area in the phase 
space plane used for the extraction with stable betatron motion inside, 
where usually the particles of the whole beam fit in at the beginning of the 
extraction. 
The extraction is executed by subsequent feeding the resonance which means that 
the particles are successively pushed out of that stable phase space area such 
that their betatron motion becomes unstable, they leave the beam, and become 
extracted by reaching the extraction septum.  
The extracted beam is also referred to as spill. 
Typical extraction time intervals reach from some hundred milliseconds up to 
hours. 
Several resonance feeding techniques are presently in operation such as 
tune sweep, betatron core driven, rf knock-out extraction, where the first two 
mechanisms are based on the reduction of the size of the stable phase space area, 
whereas the betatron amplitudes of the particles are slowly increased beyond the 
edge of the stable phase space area in the last method. 
The choice of the resonance feeding mechanism depends on the parameters of the 
beams required by the users such as beam energy and intensity, spill duration, 
or safety requirements like the ability to suddenly interrupt the extraction if the 
spill is utilised for the irradiation of patients in medical facilities. 

A frequent user requirement to the spill is a low level of undesired 
temporal structures. 
Ripples and noise on the current supplied to the accelerator quadrupole magnets 
could be identified as major source of fluctuations of the machine tune which 
translate in a ripple of the size of the stable area in the phase space 
plane and, hence, in structures on the particle flow 
out of the stable phase space area. 
These spill micro structures have a characteristic time from $\sim 10 \ {\rm ms}$ 
down to $\sim 10 \ {\rm \mu s}$ which is too short to be efficiently mitigated 
by a feedback system 
\cite{ckrantz_ipac2018_tupal036,naito_corr_of_betatron_tune_ripples_prab22_2019} 
and significantly longer than the revolution time. 

The particles need after leaving the stable phase space area a certain 
time $T_{\rm tr}$ for the transit to the extraction septum. 
These transit times are not uniform and follow a distribution instead. 
The width of the distribution defines a transit time spread $\Delta T_{\rm tr}$. 
Thereby, temporal structures formed by particles which left the stable phase 
space area at the same instant are smeared out. 
That mechanism can be utilised for reduction, especially, of spill micro structures 
while applying an extraction technique which does not rely on longitudinal dynamics 
as for example slow extraction of unbunched beam by tune sweep. 
On the other hand, 
spill micro structures can be mitigated with help of longitudinal dynamics, 
where the basic mechanism consists in the change of the momenta and, thus, 
the chromatic tunes of particles towards the resonance tune by rf fields 
which results in a transition across the edge of the stable phase space area 
much faster than that caused by the resonance feeding. 
Thereby, the duration of stay of particles near the separatrix is reduced and the 
level of structures imprinted on the particle flux out of the stable phase space 
area is reduced. 
The particle momenta can be quickly changed by applying rf fields 
of defined harmonic number 
resulting in synchrotron motion during slow extraction of bunched beams or 
channelling between empty rf buckets, or by driving longitudinal 
diffusion by rf noise with a band limited power spectrum during stochastic 
extraction. 
The aim of the present report is to study mitigation and also formation 
of of spill structures during slow extraction of bunched ion beams out of the present 
heavy ion synchrotron SIS18 of the GSI facility within a simultation study 
for synchrotron motion significantly faster than usually applied in present operation. 
Hence, the work is restricted to slow extraction by tune sweep performed 
with two fast ramped quadrupoles which is the standard slow extraction technique of SIS18. 
That topic is motivated by the plan to install a special high frequency 
rf cavity in order to provide rf fields with a harmonic number increased by an order of 
magnitude which is required by an experiment collaboration with the goal 
to take advantage of spill micro structure reduction provided by bunched 
beam extraction and to avoid spill structures at typical present rf frequencies. 
That requirement implies a strongly increased synchrotron tune. 

Generally, the results of the study show that a gain in spill quality by 
increasing the synchrotron tune and according reduction of the duration of 
stay of the particles near the oscillating edge of the stable phase space area 
can be achieved only up to an optimum of the synchrotron tune, 
whereas further rise reduces the spill quality. 
Reasons are that the transit time distribution during bunched beam slow extraction 
has a reduced but still significant influence on the spill 
quality and, in addition, is itself modified by the synchrotron motion. 
Furthermore, the creation of macroscopic spill structures is found for a 
sufficiently large synchrotron tune. 
The conditions resulting in such phenomena are analysed and possible ways of 
mitigation are proposed. 

\section{Conditions for slow extraction from SIS18 assumed in the study}

The present study is focused on the present GSI heavy ion synchrotron SIS18, 
but the results are applicable to other machines like SIS100 or hadron 
therapy synchrotrons. 

As in most synchrotrons, slow extraction from 
SIS18 is based on the excitation of a 
one-dimensional third integer resonance with sextupoles resulting in the 
formation of a triangular area of stable betatron motion in the phase space  
plane used for the extraction which is the horizontal phase plane in SIS18. 
The stable phase space area formation can be described by the
Kobayashi theory \cite{kobayashi}. 
The size of the stable phase space area is given 
by its corners which are unstable fixed points (UFP) of betatron motion. 
Written in normalised coordinates of the phase space vector 
$\vec{X}=(X,X^{'})$ with 
\begin{equation}
X \equiv \frac{x}{\sqrt{\beta_{x}}} \ \mbox{and} \ X^{'} \equiv 
\sqrt{\beta_{x}} x^{'} + \frac{\alpha_{x}}{\sqrt{\beta_{x}}} x,  
\end{equation}
the UFPs have all the same absolute value \cite{whardt_ultraslow_ex} 
\begin{equation} \label{eq_abs_xufp}
\left| \vec{X}_{\rm UFP} \right| = 
8 \pi \left| \frac{Q_{\rm r} - Q_{\rm p}}{S_{\rm v}} 
\right|, 
\end{equation}
where $Q_{\rm r}$ and $Q_{\rm p}$ are resonance tune and on axis tune of the 
particles. 
The latter can be modified by chromaticity $\xi$ and relative momentum deviation 
$\delta$ of the particles to
\begin{equation} 
Q_{\rm p}=Q_{\rm m}+\xi \delta. 
\end{equation}
$S_{\rm v}$ in Equation (\ref{eq_abs_xufp}) is the strength of a virtual sextupole 
and given by
\begin{equation} \label{eq_s_virt}
S_{\rm v} \cdot {\rm e}^{3 {\rm i} \psi_{\rm v}}=\sum \limits_{n} S_{n} 
{\rm e}^{3 {\rm i} \psi_{n}}
\end{equation}
with the normalised sextupole strength of the $n$th sextupole 
\begin{equation}
S_{n}= \frac{1}{2} \beta_{x,n}^{3/2} (k_{2} L)_{n}
\end{equation}
and the betatron phase advance $\psi_{n}$ between the location of the $n$th 
sextupole and the considered location which is 
usually the entrance of the electro-static septum. 
The $\psi_{n}$ have to be determined for the third integer tune of the 
resonance $Q_{\rm r}$. 
The resulting area of the stable phase space area is 
\begin{equation}
A_{\rm stable}=3 \frac{1}{2} \ | \vec{X}_{\rm UFP} |^{2} \ \sin 120^{o} = 
\frac{\sqrt{27}}{4} \ | \vec{X}_{\rm UFP} |^{2}. 
\end{equation}
The corresponding emittance is 
\begin{equation}
\epsilon_{\rm stable}=\frac{A_{\rm stable}}{\pi}. 
\end{equation}

The dependence of $A_{\rm stable}$ on the tune can be used for extracting the beam 
by moving the machine tune $Q_{\rm m}$ slowly across the resonance tune by changing 
the focusing strengths of quadrupoles such that it shrinks 
and the particle become successively unstable according to their transverse 
emittance and momentum dependent tune such that they leave the synchrotron. 
This technique referred to as quadrupole driven extraction is the present 
standard technique for slow extraction from SIS18. 
Hence, this manuscript is focused on the usage of this method. 

A very comprehensive series of machine experiments was performed with beams of Ar$^{18+}$ 
ion at the beam energy $E=500 \ {\rm MeV/u}$. 
Typical particle numbers were between $10^{5}$ and $10^{6}$ ions per cycle 
which is lower than in user experiments. 
The purpose was to use a scintillation counter for measuring the spill 
which has a limited count rate. 
But it allows for experiments with a high time resolution which was 
actually chosen to $10 \ {\rm \mu s}$. 
In addition, it provides the opportunity to perform particle tracking simulations 
with particle numbers similar to those of the measurements. 
These experiments delivered the most important results the particle tracking results 
achieved in this study are compared with. 
For that reason, we apply conditions in the simulation of the present study similar to 
those in these experiments. 

\section{Spill simulations and characterisation of spill micro structures}

Spill structures can be investigated by simulating the 
process of slow extraction with particle tracking. 
We use the MAD-X code for the simulations. 
Gaussian distribution functions truncated at $2 \sigma$ are applied to set 
the initial coordinates of the test particles. 
$10^{5}$ particles were tracked in each simulation for $5 \cdot 10^{5}$ revolutions 
after finishing the bunching process. 
The $\sigma$ values of corresponding Gaussian distribution functions of 
the transverse particle coordinates 
are defined by the rms beam emittances $\epsilon_{x,{\rm rms}}=3.9 \ {\rm mm \ mrad}$ 
and $\epsilon_{y,{\rm rms}}=1.3 \ {\rm mm \ mrad}$. 
The resulting maximum beam emittances are 
$\epsilon_{x,{\rm max}}=15.5 \ {\rm mm \ mrad}$ and 
$\epsilon_{y,{\rm max}}=5.2 \ {\rm mm \ mrad}$, respectively. 
They arise from the beam width given by the machine acceptance during injection 
and subsequent adiabatic shrinkage during acceleration. 
In order to study the influence of the horizontal beam width, 
transverse beam emittances reduced by the factor $0.2$ are applied. 
The maximum momentum deviations $\delta_{\rm m}$ reach from 
$\delta_{\rm m}=5 \cdot 10^{-4}$ up to $\delta_{\rm m}=10^{-3}$, 
where the actual value depends on the rf voltage defined by amplitude $V_{\rm rf}$ 
and harmonic number $h$. 

The time resolution for measuring spills is given by the measurement rate of 
the spill measurement system, where $f_{\rm sample}=100 \ {\rm kHz}$ in SIS18. 
The resulting upper limit to the frequency range of the spill micro structures 
is then $f_{\rm up}=f_{\rm sample}/2=50 \ {\rm kHz}$. 

The quantity used for characterising the spill quality defined by the 
micro structure level in this study is the time dependent duty factor 
\cite{rsingh_ipac2018} 
\begin{equation}
F(t) = \frac{\langle N \rangle^{2}(t)}{\langle N^{2} \rangle(t)}
=\frac{N_{\rm av}^{2}(t)}{N_{\rm av}^{2}(t)+\sigma_{N}^{2}(t)} 
\end{equation}
is applied, where $\langle x \rangle$ denotes the average of the variable $x$ 
given in time intervals according to the measurement resolution 
$t_{\rm meas}=1/f_{\rm meas}$ in time intervals $t_{\rm rec}$ which define the 
average length. 
The latter defines the recording resolution. 
The actual size of the measurement and recording time intervals 
used for this study were $t_{\rm meas}=10 \ {\rm \mu s}$ 
and $t_{\rm rec}=10 \ {\rm ms}$. 
A good spill quality given by a low spill structure level results in large 
value of the time dependent duty factor. 

The upper limit to $F(t)$ under realistic beam conditions and for 
$t_{\rm rec} \gg t_{\rm meas}$ is reached in the 
Poisson limit which denotes stochastic distribution of extracted 
particles in the time intervals. 
For Poisson's distribution function to find $N$ particles in a measurement 
time interval, 
\begin{equation}
f_{\rm P}={\rm e}^{-\lambda} \frac{\lambda^{N}}{N!}
\end{equation}
one finds $N_{\rm av}^{2} + \sigma_{N}^{2}=(N^{2})_{\rm av}=N_{\rm av} 
\left( N_{\rm av} + 1 \right)$ such that 
\begin{equation}
F_{\rm P}(t)=\frac{N_{\rm av}(t)}{N_{\rm av}(t)+1}. 
\end{equation}

It can be useful to reduce the time dependent duty factor to the weighted 
factor defined in \cite{rsingh_phys_rev_applied} by 
\begin{eqnarray} \label{eq_av_duty_factor}
F_{\rm av}&=&\frac{\int {\rm d} t \ \dot{N}(t) F(t)}{\int {\rm d} t \ \dot{N}(t)} 
\nonumber \\
&=& \frac{\sum \limits_{k} N_{k} F_{k}}{\sum \limits_{k} N_{k}} 
\end{eqnarray}
which is, basically, the time average of the weight function $\dot{N}(t)$ which is 
the extraction rate at instant $t$ denoting the spill. 
On the other hand, spill and time dependent duty factor are recorded 
with the time resolution given by $t_{\rm rec}$ 
such that, approximately, $N_{k}=\dot{N}(t_{k}) \cdot t_{\rm rec}$ with 
$t_{k} = k \cdot t_{\rm rec}$ and 
the integrals represented by the sums in the second row 
of Equation (\ref{eq_av_duty_factor}). 

\section{Mitigation of spill micro structures with bunching} 
\label{sec_mitigation_micro_structure}

The spread of the particles' transit times $\Delta T_{\rm tr}$ arising from the 
width of the transit time distribution results in a finite interval of arrival times 
of particles which became unstable at the same instant. 
Hence, spill micro structures imprinted by oscillations of the stable 
phase space area's size due to a tune ripple are smeared out. 
This mechanism could be identified as a major spill smoothing 
mechanism for slow extraction of coasting beams 
\cite{ssorge_ipac2018,rsingh_phys_rev_applied}. 

\begin{figure}
\centerline{
\includegraphics*[width=100mm]{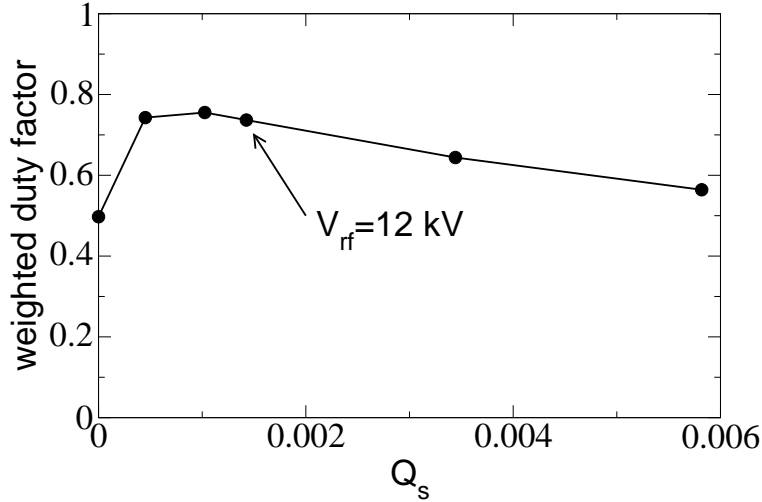}
}
\caption{Weighted duty factors according to Equation (\ref{eq_av_duty_factor}) 
of spills obtained in simulations of the slow extraction of Ar$^{18+}$ 
beams from SIS18 at the beam energy $E=500 \ {\rm MeV/u}$ 
as function of the synchrotron tune according to the 
the rf voltages $V_{\rm rf}=(0, \ 1.2, \ 6.2, \ 12, \ 70, \ 200) \ {\rm kV}$, 
where $V_{\rm rf}=12 \ {\rm kV}$ is the maximum which can be reached in SIS18. 
The rf voltage at the beginning of the simulations is switched off and 
slowly turned on during the first 50000 turns. 
The initial maximum momentum width is $\delta_{\rm m}=5 \cdot 10^{-4}$ and is 
increased during the bunching process. 
Hence, also longitudinal dynamics with low rf voltages with particles captured and 
others not captured in the rf bucket are well described. 
}
\label{fig_av_duty_factor_ar18_ekin0.5gev_k2la0.05_h5_all_vrf}
\end{figure}

In contrast to the aforementioned mechanism, 
spill smoothing for bunched beam slow extraction is mainly based on the periodic 
change of the particles' tune due to synchrotron motion which yields a transition 
from stable to unstable betatron motion significantly faster than caused by the tune sweep. 
Therefore, the particles will stay much shorter near the oscillating edge of 
the stable phase space area such that resulting structures imprinted on the 
spill are strongly reduced. 
Following this reasoning, 
one would expect that the spill quality defining duty factor always 
grows for an increasing synchrotron tune $Q_{\rm s}$ or, at least, saturates because 
the time the particles are near the separatrix can not be reduced below zero. 
However, it is found by performing particle tracking simulations of the 
extraction that the weighted duty factor has a maximum 
at a certain $Q_{\rm s}$ and is decreased if $Q_{\rm s}$ is further increased 
which one can see in Figure 
\ref{fig_av_duty_factor_ar18_ekin0.5gev_k2la0.05_h5_all_vrf}. 
Our analysis reveals that particles are re-captured into the stable phase space area 
after becoming unstable if the synchrotron motion is too fast such that they can 
not become extracted. 
As a result, the distribution of the transit times of extracted particles is altered. 
That suggests that the influence of the transit time distribution is only reduced but not 
irrelevant. 

\begin{figure}
\centerline{\includegraphics*[width=100mm]{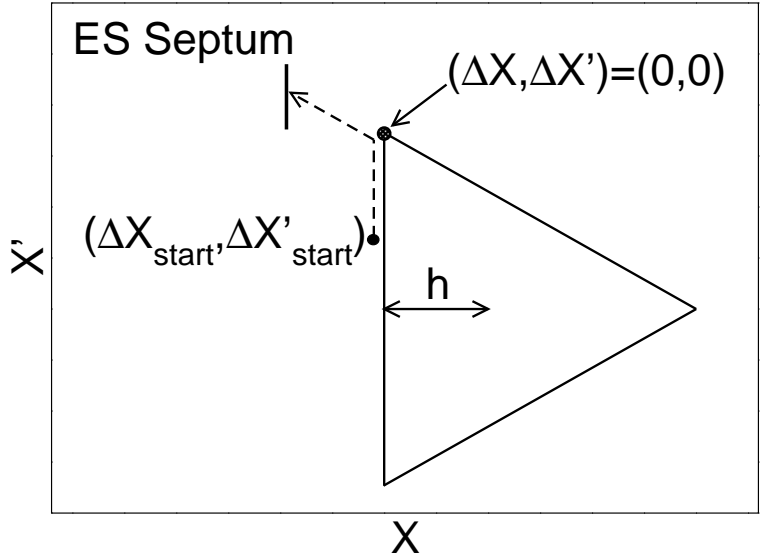}}
\caption{Schematic representation of the stable phase area in normalised 
coordinates $X,X^{'}$ defined relative to he centre of the triangle. 
The starting coordinates $\Delta X_{\rm start}$, $\Delta X_{\rm start}^{'}$ 
relevant for the time of a particle's transit from the stable phase area to the 
electro-static septum are defined relative to the upper left corner of the 
triangle. }
\label{fig_separatrix_schematic_pimms_nom}
\end{figure}

We apply a model based on the analytic expression for the transit time from 
Equation (4.13) in \cite{pimms}
\begin{equation} \label{eq_t_tr_pimms4.13}
T_{\rm tr} \approx \frac{1}{\sqrt{3} \varepsilon} \ln \left| 
\frac{1}{1 - \Lambda_{0}^{2}} \frac{n}{n+3} 
\frac{3}{\lambda_{0}} \right|, 
\end{equation}
in order to demonstrate the effect of the synchrotron motion on the transit time 
distribution. 
The variables in this equation are 
$\varepsilon=6 \pi | Q_{x,{\rm r}}-Q_{x}(\delta) |$, 
$\lambda_{0} \equiv - \Delta X_{\rm start}/h$ and 
$\Lambda_{0} \equiv - \Delta X_{\rm start}^{'}/(2 \sqrt{3} h)$ 
with the the coordinates of the start 
of a particles transition in horizontal phase space 
$\Delta X_{\rm start},\Delta X_{\rm start}^{'}$, 
and $n \equiv - \Delta X_{\rm ESS}/h$. 
One should note that $\Delta X_{\rm start}$, $\Delta X_{\rm start}^{'}$, and the 
normalised horizontal coordinate of the electro-static septum $\Delta X_{\rm ESS}$ 
are defined relative to the point $(\Delta X, \Delta X^{'})=(0,0)$ at the upper 
left corner of the triangular stable phase space area shown in Figure 
\ref{fig_separatrix_schematic_pimms_nom} and, hence, negative. 
The transit time in this formula is dimensionless and given in units of three 
subsequent revolutions. 
By applying the analytic formula, the effect can be directly depicted. 
That is an advantage compared to the use of simulation data, where it is 
difficult to determine the instant when a particle becomes unstable and 
its transit from the stable phase area to the extraction septum begins. 

\begin{figure}
\centerline{\includegraphics*[width=90mm]{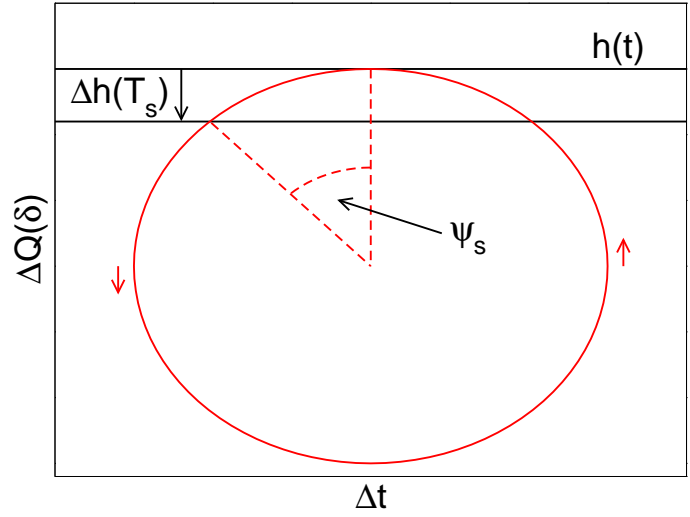}}
\caption{Range of longitudinal coordinates of particles with certain horizontal 
emittance related to the size of the stable phase space area characterised by 
$h(t)$ at instant $t$ and shrunk within a 
synchrotron period $T_{\rm s}$ by $\Delta h(T_{\rm s})$. }
\label{fig_bunch_with_stability_frontier_delta_t_vs_delta_qx}
\end{figure}

From Equation (\ref{eq_t_tr_pimms4.13}), it follows that the transit time of a 
particle is the shorter the larger $\lambda_{0}$ is, where a large value of $\lambda_{0}$ 
corresponds to a large distance of a particles'  
starting point $\Delta X_{\rm start}$ from the separatrix. 
The maximum distance $\Delta X_{\rm start}$ 
during a synchrotron period is reach at the cusp of synchrotron 
motion, where the momentum dependent tune 
\begin{equation}
Q(\delta)=Q_{\rm m}+\Delta Q(\delta) \ \mbox{with} \ \Delta Q(\delta) = \xi \delta
\end{equation}
is closest to the resonance tune $Q_{\rm r}$. 
On the other hand, the chromatic tune of particle will move away from the 
resonance after it has passed the cusp and after a certain time it will 
re-enter the stable phase space area if it has not reached the extraction 
septum, yet. 
This limiting time $t_{\rm lim}$ can approximately be defined by
\begin{equation}
t_{\rm lim}=T_{\rm s} \frac{\psi_{\rm s}}{2 \pi}, 
\end{equation}
where $\psi_{\rm s}$ is the synchrotron phase angle shown in Figure 
\ref{fig_bunch_with_stability_frontier_delta_t_vs_delta_qx}. 
The ellipse in this figure represents the trajectory in 
longitudinal phase space with the largest synchrotron amplitude which 
corresponds to the maximum $\psi_{\rm s}$. 
$\psi_{\rm s}$ will be the smaller the 
less the cusp of the trajectory goes beyond the stability frontier at the instant 
$t+T_{\rm s}$. 
In addition, particles with a short $t_{\rm lim}$ will start their transit 
from only a little outside the stable phase space area at $t+T_{\rm s}$. 
Hence, $\Delta X_{\rm start}$ and $\lambda_{0}$ of such particles in Equation 
(\ref{eq_t_tr_pimms4.13}) is small and will result in a long transit time. 
In other words, longer transit times $T_{\rm tr}$ will be suppressed by shorter 
limiting times $t_{\rm lim}$. 

In order to demonstrate the effect, the transit time distribution will be 
determined in two steps for conditions of an 
Ar$^{18+}$ beam at the beam energy $E=500 \ {\rm MeV/u}$. 
The two rf voltages $V_{\rm rf}=6.2 \ {\rm kV}$ and $V_{\rm rf}=12.0 \ {\rm kV}$ 
are applied which are the largest used in the measurements. 
Further simplifying assumptions are: 
\begin{enumerate}
\item the distribution of the momentum deviations $\delta$ 
is symmetric with respect to the sign of $\delta$ and the maximum for all cases is 
$\delta_{\rm m}=0.0005$, 
\item the synchrotron tune does not depend on the synchrotron amplitude 
and is given by \cite{sylee} 
\begin{equation}
Q_{\rm s}=
\sqrt{ \frac{Z e h V_{\rm rf} |\eta \cos \phi_{\rm s}|}{2 \pi \beta^{2} E_{\rm tot}} }
\end{equation}
with the charge of the ions $Z e$, harmonic number and amplitude of the rf 
voltage $h,V_{\rm rf}$, the phase slip factor $\eta$, the phase of the synchronous 
particle $\phi_{\rm s}$, the relativistic factor $\beta=v/c$, and the total 
energy of an ion $E_{\rm tot}=A m_{\rm u} c^{2} \gamma$ with the number of nucleons 
in the ion's nucleus, $m_{\rm u} c^{2}=931.494 \ {\rm MeV}$ and 
$\gamma=1/\sqrt{1-\beta^{2}}$. 
\item the particle density does not depend on betatron and synchrotron 
coordinates. 
\end{enumerate}

\begin{figure}
\centerline{\includegraphics*[width=110mm]{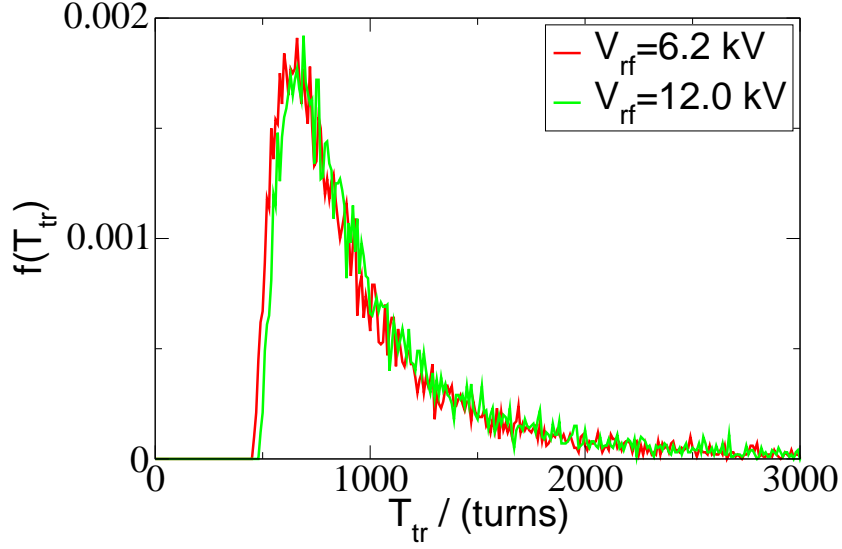}}
\caption{Distributions of the transit times according to Equation 
(\ref{eq_t_tr_pimms4.13}) without taking into account the 
limitation arising from the re-capture of particles into the stable phase area 
due to synchrotron motion for the same conditions as those for Figure 
\ref{fig_av_duty_factor_ar18_ekin0.5gev_k2la0.05_h5_all_vrf}. }
\label{fig_t_transit_1st_tsynch_vrf}
\end{figure}

In the first step, the transit time distribution is determined according to 
the initial particles $\Delta X_{\rm start}, \Delta X_{\rm start}^{'}$ 
as well as the momentum dependent difference between particle and resonance 
tunes but neglecting the limitation of the transit times due to the synchrotron 
motion. 
The resulting transit time distributions are shown in Figure 
\ref{fig_t_transit_1st_tsynch_vrf}. 
They appear rather independent from the synchrotron tune. 

\begin{figure}
\centerline{\includegraphics*[width=110mm]{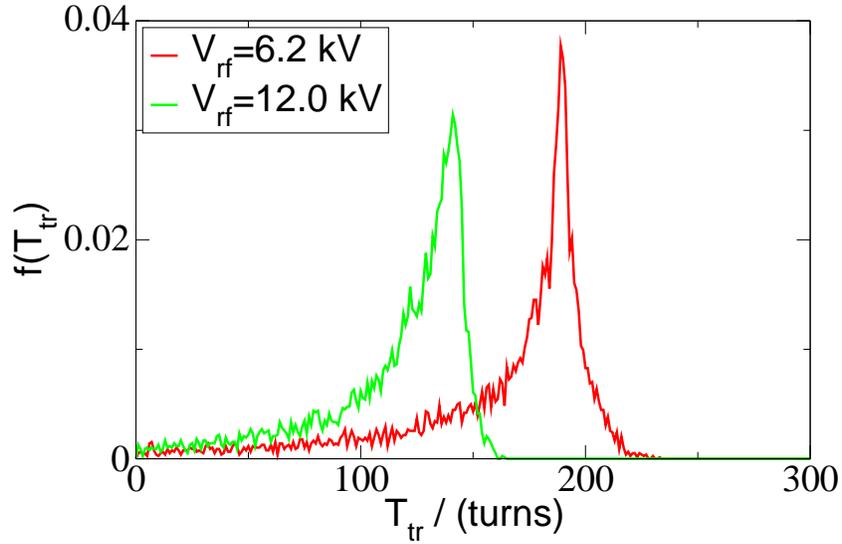}}
\caption{Distributions of the transit times as determined in Figure 
\ref{fig_t_transit_1st_tsynch_vrf}, 
where the limitation is taken into account 
which arises from the re-capture of particles into the stable phase area 
due to synchrotron motion. 
Hence, the distributions in this figure represent the real transit times of the 
particles which reach the extraction septum. }
\label{fig_t_transit_arrived_vrf}
\end{figure}

In the second step, the limitation of the transit times due to the synchrotron 
motion is taken into account. 
It turns out that, after a single synchrotron period, none of the particles 
is extracted for all considered rf voltages. 
For that reason, the machine tune is further moved towards the resonance 
in time steps of the synchrotron period. 
In addition, we assume that the betatron amplitude of the particles has not changed during 
the short stay outside the stable phase space area such that the horizontal start coordinates 
relative to the centre of the stable phase area, $X_{\rm start}, X_{\rm start}^{'}$ are not changed. 
The distance of the starting point from the separatrix, $\Delta X_{\rm start}$ and, hence, 
$\lambda_{0}$, of all particles will increase because of the shrinking stable phase area denoted by a 
decreasing $h(t)$ in the transit time formula above. 
That will result in a decrease of the transit times $T_{\rm tr}$ 
such that they will become shorter than the limiting times $t_{\rm lim}$. 
The resulting in the transit time distributions are those of the particles 
which reach the extraction septum. 
They are shown in Figure \ref{fig_t_transit_arrived_vrf}. 
\begin{figure}
\centerline{\includegraphics*[width=90mm]{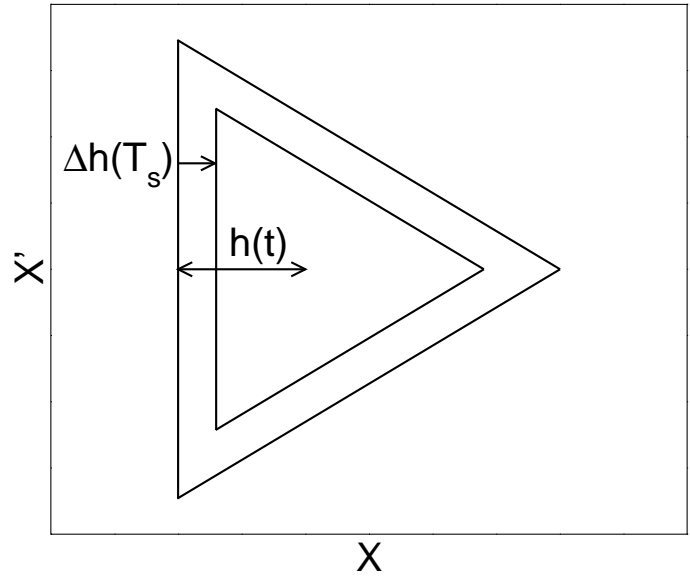}}
\caption{Schematic plot of the stable phase space area at instant $t$ and shrunk 
within a synchrotron period by moving the machine tune $Q_{x,{\rm m}}$ towards the 
resonance tune $Q_{x,{\rm r}}$.}
\label{fig_shrunk_separatrix_schematic}
\end{figure}
The corresponding averaged transit times and rms spreads are 
\begin{eqnarray}
V_{\rm rf}=6.2 \ {\rm kV} &:& T_{\rm tr,av}=162 \ {\rm rev.}, \ 
\Delta T_{\rm tr,rms}=44 \ {\rm rev.} \nonumber \\
V_{\rm rf}=12.0 \ {\rm kV} &:& T_{\rm tr,av}=116 \ {\rm rev.}, \ 
\Delta T_{\rm tr,rms}=33 \ {\rm rev.} \nonumber 
\end{eqnarray}
The results suggest that the decrease of the transit time spread for 
the rf voltage increased from $6.2 \ {\rm kV}$ to $12 \ {\rm kV}$ which results in 
the decrease of the duty factor has a {\bf strong}er impact than the 
decrease of the stay of the particles near the oscillating edge of the horizontal 
stable phase space which would result in a reduced spill micro structure level. 

One should note that the differences of the mean and rms values of the 
transit times found for the considered rf voltages are rather underestimated 
because the increase of the synchrotron period for increased synchrotron 
amplitude is neglected which rather occurs for lower voltages. 

\section{Spill nano structures}

The improvement of the spill quality on microscopic time scales due to 
bunching occurs at the price of the creation of spill structures 
on time scales defined by the rf frequency
\begin{equation}
f_{\rm rf}=f_{\rm rev} \cdot h 
\end{equation} 
with the revolution time $f_{\rm rev}$ and the harmonic number of the rf fields $h$. 
The rf frequency of the present SIS18 rf cavity reaches up to 
$f_{\rm rf}=6 \ {\rm MHz}$ \cite{sis18_param_list}. 
According to a typical revolution time $t_{\rm rev} \in [0.8,1.0] \ {\rm \mu s}$ 
during extraction, the rf harmonic number is usually $h=4$ or $h=5$ and the 
total duration of an rf bucket is $t_{\rm bucket} \in [200,250] \ {\rm ns}$. 
Hence, these structure are referred to as spill nano structures. 

The length of the bunches in the extracted beam is found to be shorter than that 
in the circulating beam \cite{tmilosic_wepp28_ibic2021_pohang}. 
The reason is that only particles are extracted which are near the cusp of 
synchrotron motion such that their chromatic tune reaches the 
phase space range of unstable betatron motion which is, basically, 
the area in the longitudinal trajectory marked by the red ellipse in Figure 
\ref{fig_bunch_with_stability_frontier_delta_t_vs_delta_qx} and the black solid lines 
which mark the frontier between stable and unstable motion at beginning and 
ending of a synchrotron period. 
In addition, most particles will start with their transit towards the extraction 
channel after they have passed the cusp of synchrotron motion because the transit time 
reaches its minimum there which is pointed out in Section 
\ref{sec_mitigation_micro_structure}. 

\begin{figure}

\centerline{
\includegraphics*[width=100mm]{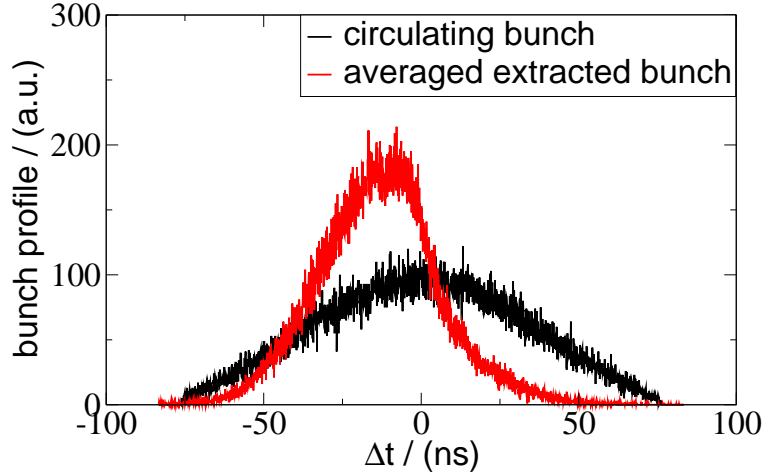}
}
\caption{Intensity profiles as function of the time difference to 
the synchronous particle of a circulating bunch and the resulting extracted 
bunches averaged over the time of the whole extraction process obtained in a simulation 
for conditions used in Figure \ref{fig_av_duty_factor_ar18_ekin0.5gev_k2la0.05_h5_all_vrf}
with $h=5$ and $V_{\rm rf}=6.2 \ {\rm kV}$.
}
\label{fig_bunch_profiles_circulating_vs_extracted}
\end{figure}

An example for resulting profiles of circulating and extracted bunches as 
function of the arrival time compared to that of the synchronous particle 
obtained from simulation data are shown in Figure 
\ref{fig_bunch_profiles_circulating_vs_extracted}. 
The length given by the full length at half maximum intensity of the 
circulating bunch under the chosen conditions is about 
$t_{\rm b,circ}=84 \ {\rm ns}$, whereas 
that of the extracted bunch integrated over the whole extraction process is 
$t_{\rm b,ex}=40 \ {\rm ns}$. 
The full bucket length is $t_{\rm bucket}=190 \ {\rm ns}$ such that the 
duration of the void between two bunches is $t_{\rm void,circ}=106 \ {\rm ns}$ 
for the circulating beam and even $t_{\rm void,ex}=150 \ {\rm ns}$ for the 
extracted beam. 
In addition, one should keep in mind that the extracted bunch in Figure 
\ref{fig_bunch_profiles_circulating_vs_extracted} is averaged over the whole 
extraction such that it is longer than an instantaneous bunch. 
As a result, the intensity of the beam arriving at the target will have a {\bf strong} 
variability. 
That limits the maximum intensity and, hence, the event rate of some 
experiments because the integration time of the detector is comparable 
or shorter than the duration of a bunch. 
E. g. the HADES detector has an integration time of $t_{\rm int}=140 \ {\rm ns}$. 
For that reason, the number of particles detected in an integration time 
will be given by the maximum intensity which can lead to a pile up in case of a 
too large averaged intensity or it will be zero when a void is detected which 
means waste of detection capacity 
\cite{pietraszko_slowex_darmstadt_2016,pietraszko_hic4fair_darmstadt_20160226}. 
A possible solution is to increase the rf frequency to at least 
$f_{\rm rf} \ge 40 \ {\rm MHz}$. 
Actually, the installation of an rf cavity with a frequency $f_{\rm rf}=80 \ {\rm MHz}$ 
is planned such that the duration of an rf bucket would shrink to 
$t_{\rm bucket}=12.5 \ {\rm ns}$ and always five or six bunches arrive within 
an integration time at the detector \cite{pschmid_slowex_cern2017}. 
The result would be a much lower fluctuation level of particles detected in 
subsequent detector integration times. 
The generation of bunches with such a high rf frequency results 
in {\bf fast} synchrotron motion because $Q_{\rm s} \propto \sqrt{h}$. 

\section{Macroscopic spill structures arising from {\bf fast} synchrotron motion}

\begin{figure}
\centerline{
\includegraphics*[angle=90,width=100mm]{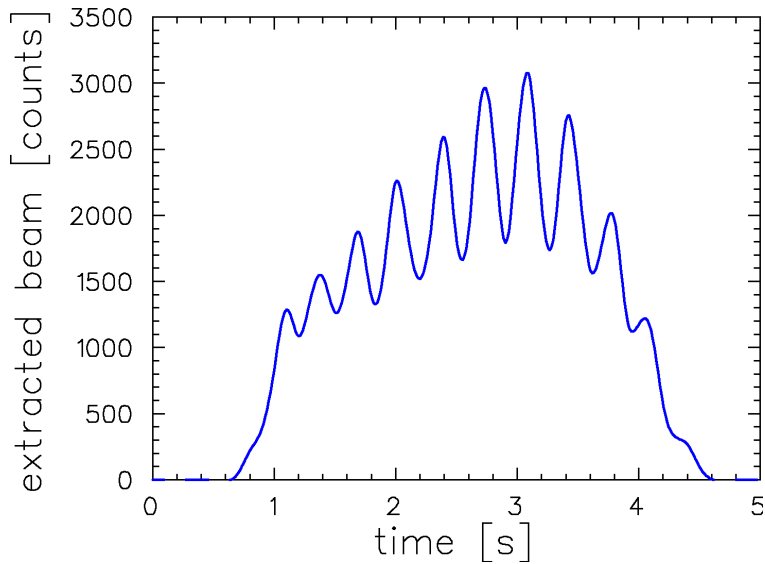}
}
\caption{Spill of an Ar$^{18+}$ beam measured 
during slow extraction from SIS18 at the beam energy $E=500 \ {\rm MeV/u}$ 
bunched with the rf voltage of amplitude $V_{\rm rf}=12 \ {\rm kV}$ and 
harmonic number $h=5$. }
\label{fig_meas_spill_ar18_ekin500mev_vrf12kv}
\end{figure}

\begin{figure}
\centerline{
\includegraphics*[width=110mm]{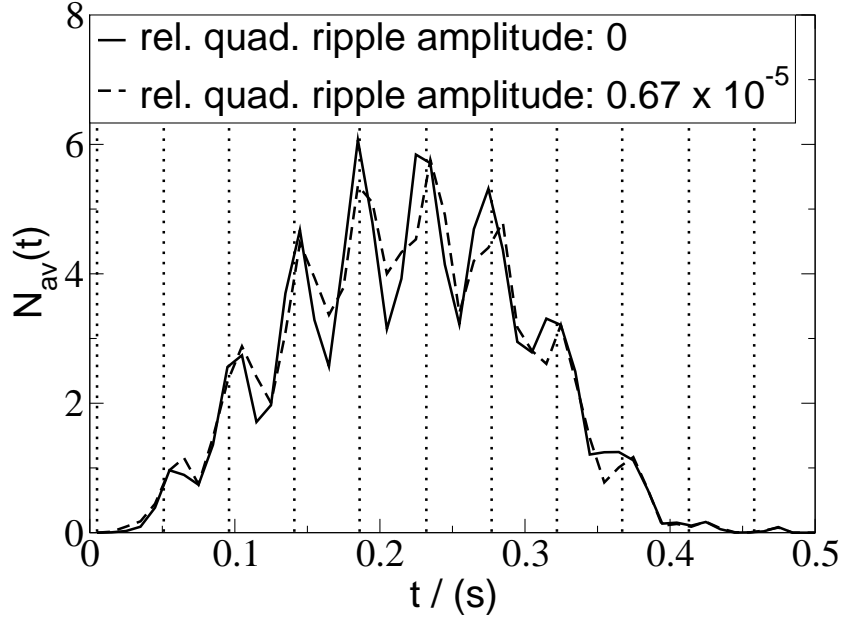}
}
\caption{Spill of an Ar$^{18+}$ beam obtained in a slow 
extraction simulation for SIS18 conditions at the beam energy 
$E=500 \ {\rm MeV/u}$, where $V_{\rm rf}=12 \ {\rm kV}$, $h=5$, 
and quadrupole ripples of two different amplitudes are applied. 
The transverse beam emittances are correspond to the machine acceptance 
at injection reduced by the factor $0.2$ and adiabatic shrinkage during 
acceleration afterwards. 
Furthermore, the maximum momentum deviation in the bunches is 
$\delta_{\rm m}=5 \cdot 10^{-4}$. 
Both, the transverse emittances and the maximum momentum deviation are 
lower than during usual SIS-18 operation. 
The dotted vertical lines mark the moments when the machine tune crosses a 
synchro-betatron resonance. 
The third integer resonance corresponding to $m=0$ in Equation 
(\ref{eq_synchro_betatron_resonance}) is crossed at $t=0.41 \ {\rm s}$. }
\label{fig_sim_spill_ar18_ekin500mev_vrf12kv}
\end{figure}

It is shown in Section \ref{sec_mitigation_micro_structure} that there is an 
optimum of the synchrotron tune above which the level of spill micro structures 
is again increased for further increased $Q_{\rm s}$. 
An additional limitation to the improvement of the spill quality found in our studies 
is that synchrotron motion with sufficiently large synchrotron tune can 
result in the formation of macroscopic spill structures of duration 
$\sim 0.1 \ {\rm s}$ or longer. 
Such structures could be found for conditions of 
present SIS18 operation in measurements as well as in simulations, see 
Figures \ref{fig_meas_spill_ar18_ekin500mev_vrf12kv} and 
\ref{fig_sim_spill_ar18_ekin500mev_vrf12kv}. 
These figures show spills of the extraction Ar$^{18+}$ beams at beam energy 
$E=500 \ {\rm MeV/u}$ bunched with an rf voltage of amplitude 
$V_{\rm rf}=12 \ {\rm kV}$ and harmonic number $h=5$. 
Crossing synchro-betatron coupling resonances could be identified 
in tracking simulations as a likely mechanism. 
These resonances are defined by the condition 
\cite{piwinski_synchro_betatron_resonances}
\begin{equation} \label{eq_synchro_betatron_resonance}
k Q_{x} + l Q_{y} + m Q_{s}=n, 
\end{equation}
where $k,l,m,n$ are integer numbers, where $k=3$, $l=0$, and 
$n=3 \cdot Q_{x,{\rm r}}=13$. 
In simulations, the exact time behaviour of $Q_{x}$ is known 
and, hence, it can be detected when such a resonance is crossed. 
In doing so, a clear correlation between the instants when $Q_{x}$ is fulfilling 
the resonance condition and the appearance of the structures could be 
found which one can see in Figure \ref{fig_sim_spill_ar18_ekin500mev_vrf12kv}. 
That figure makes also visible that quadrupole ripples inhibit the 
formation of macroscopic structures, where the spill obtained with 
the quadrupole ripple of amplitude zero resembles more the measured 
spill in Figure \ref{fig_meas_spill_ar18_ekin500mev_vrf12kv}. 

\begin{figure}
\centerline{
\includegraphics[width=110mm]{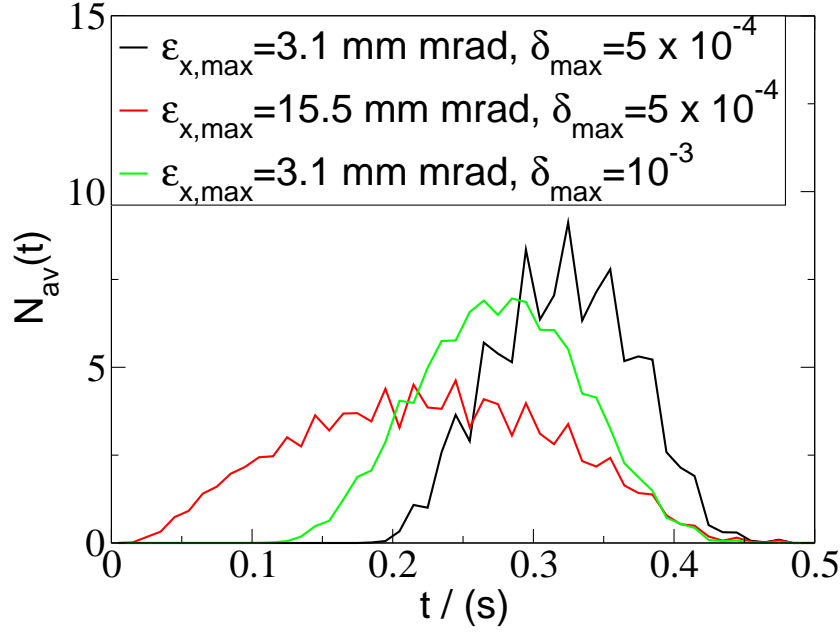}
}
\caption{
Spills obtained for Ar$^{18+}$ beam at beam energy $E=500 \ {\rm MeV/u}$ 
bunched with the rf voltage of harmonic number $h=5$ and amplitude 
$V_{\rm rf}=12 \ {\rm kV}$ with varied horizontal emittance and momentum width. 
The black curve corresponds to the solid black curve in Figure 
\ref{fig_sim_spill_ar18_ekin500mev_vrf12kv}, where the initial horizontal tune 
is changed from $Q_{x,{\rm ini}}=4.329$ in the figure above to 
$Q_{x,{\rm ini}}=4.327$ in this figure in order to enable the application of increased 
horizontal emittance and momentum width. 
}
\label{fig_sim_spill_ar18_ekin500mev_vrf12kv_varied_epsxmax_dpmax}
\end{figure}

The macroscopic structures are found to depend on 
the rf voltage, the horizontal beam emittance, and the momentum width, where 
the macroscopic spill structures become more visible for a higher rf voltage 
because the distance between the resulting synchro-betatron resonances 
is larger such that they can be better resolved. 
In addition, the structures shown in Figure 
\ref{fig_sim_spill_ar18_ekin500mev_vrf12kv} are found 
for bunches with a maximum momentum deviation $\delta_{\rm max}=5 \cdot 10^{-4}$ 
and a horizontal beam emittance $\varepsilon_{x}=3.1 \ {\rm mm \ mrad}$. 
The latter is achieved by filling the total machine acceptance 
at injection to $20 \ \%$ and subsequent adiabatic shrinkage 
during acceleration. 
Both values are lower than during regular SIS18 operation. 
The structures are diminished or even vanish if the horizontal emittance or 
the momentum width is increased. 
It turns out that the horizontal emittance has a {\bf strong}er influence than the momentum 
width which is demonstrated by repeating the simulations 
with increased momentum width and horizontal emittance, respectively. 
The resulting spills are shown in Figure 
\ref{fig_sim_spill_ar18_ekin500mev_vrf12kv_varied_epsxmax_dpmax}. 
They suggest that macroscopic spill structures should not play an 
important role in present SIS18 operation. 

\begin{figure}
\centerline{
\includegraphics*[width=110mm]{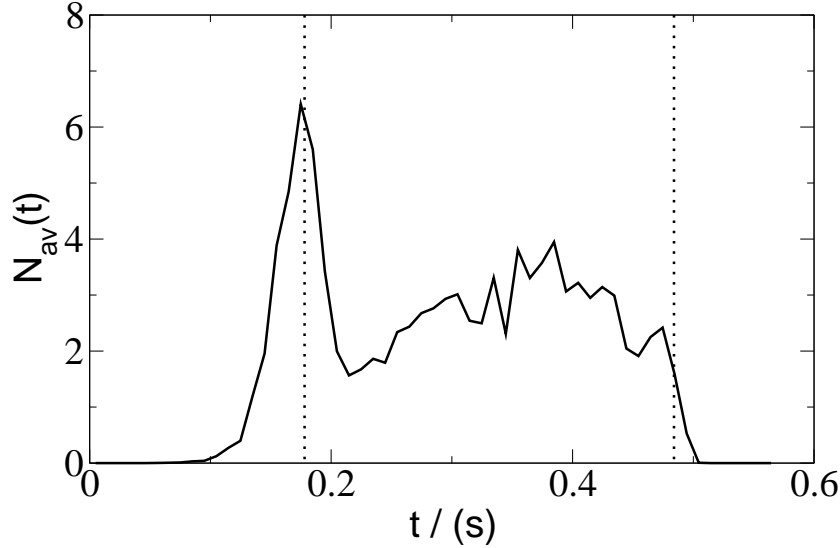}
}
\caption{Spill of the extraction of an Ar$^{18+}$ beam at beam energy 
$E=500 \ {\rm MeV/u}$ bunched with rf fields of the voltage $V_{\rm rf}=70 \ {\rm kV}$ 
and the harmonic number $h=100$ obtained by particle tracking. 
The horizontal tune is moved from $Q_{x,{\rm i}}=4.326$ to $Q_{x,{\rm f}}=4.334$. 
Two synchrotron-betatron resonances are crossed which are marked by the dotted 
vertical lines. }
\label{fig_sim_spill_ar18_500mev_vrf70kv_h100}
\end{figure}

The situation will be different if the harmonic number of the rf field $h$ is 
increased by a factor of about $20$ which will result in a {\bf strong}ly increased 
synchrotron tune. 
That is demonstrated in the following by examining the example of an 
Ar$^{18+}$ beam extracted at the same energy as in the example above but the 
harmonic number increased to $h=100$. 
In the first step, the rf voltage amplitude to $V_{\rm rf}=70 \ {\rm kV}$ 
is assumed which is planned to be the maximum 
of the high frequency cavity foreseen to be installed in SIS18 
\cite{pschmid_slowex_cern2017}. 
The resulting spill obtained by particle tracking is 
shown in Figure \ref{fig_sim_spill_ar18_500mev_vrf70kv_h100}. 
The horizontal tune in this example is shifted from $Q_{x,{\rm i}}=4.326$ to 
$Q_{x,{\rm f}}=4.334$. 
The synchrotron tune is $Q_{\rm s}=0.015$ such that the gap between two 
synchro-betatron resonances is a little lower than three times the horizontal tune 
interval crossed during the extraction, $\Delta Q_{x}=0.024$. 
Two resonance are crossed which are marked by the 
dotted lines in Figure \ref{fig_sim_spill_ar18_500mev_vrf70kv_h100}, 
where the horizontal third integer resonance is that crossed at $t=0.48 \ {\rm s}$. 
In addition, this figure shows that the extraction rate has its maximum near the 
resonance crossed first at $t=0.18 \ {\rm s}$. 
The reason is that the number of particles in the beam when reaching the 
third integer resonance is already {\bf strong}ly reduced. 

\begin{figure}
\centerline{
\includegraphics*[width=110mm]{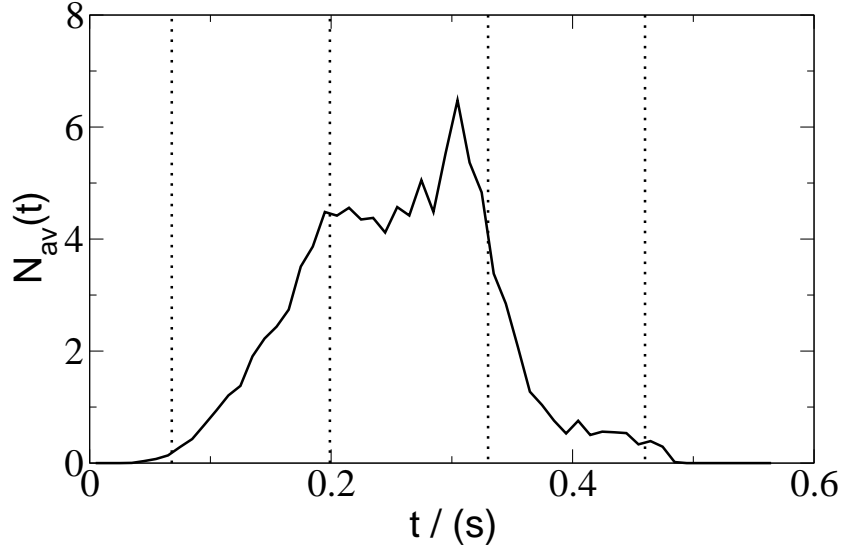}
}
\caption{Spill of the extraction of an Ar$^{18+}$ beam at beam energy 
$E=500 \ {\rm MeV/u}$ bunched with rf fields of the voltage $V_{\rm rf}=5 \ {\rm kV}$ 
and the harmonic number $h=100$ obtained by particle tracking. 
The horizontal tune is moved from $Q_{x,{\rm i}}=4.329$ to $Q_{x,{\rm f}}=4.334$. 
The dotted vertical lines denote the instants when synchro-betatron resonances 
are crossed, the dashed vertical line at $t=0.048 \ {\rm s}$ denotes the 
ending of the rf voltage ramp, and the dashed line at $t=0.38 \ {\rm s}$ 
denotes the crossing of the third integer resonance. }
\label{fig_sim_spill_ar18_500mev_vrf5kv_h100}
\end{figure}

A significant reduction of the influence of the synchro-betatron resonance crossing 
occurs for reduced rf voltage which is demonstrated in the second step. 
Figure \ref{fig_sim_spill_ar18_500mev_vrf5kv_h100} shows a spill 
obtained in a simulation, where the rf voltage is reduced to $V_{\rm rf}=5 \ {\rm kV}$ 
and the horizontal tune interval is changed to 
$(Q_{x,{\rm i}},Q_{x,{\rm f}})=(4.329,4.334)$. 
All other quantities are preserved from the previous example. 
Hence, the synchrotron tune is $Q_{\rm s}=0.0041$ and three times the horizontal 
tune interval is $3 \cdot (Q_{x,{\rm f}} - Q_{x,{\rm i}})=0.015$ such that 
four resonances are crossed, where the last is the horizontal third integer resonance. 
The figure shows that the maximum extraction rate occurs between crossing the 
second and the third crossed resonance. 
That is a sign that both these resonances are smeared out because they are closer to 
each other than for $V_{\rm rf}=70 \ {\rm kV}$ 
and because the synchrotron tune varies with the synchrotron amplitude. 

\begin{figure}
\centerline{
\includegraphics*[width=110mm]{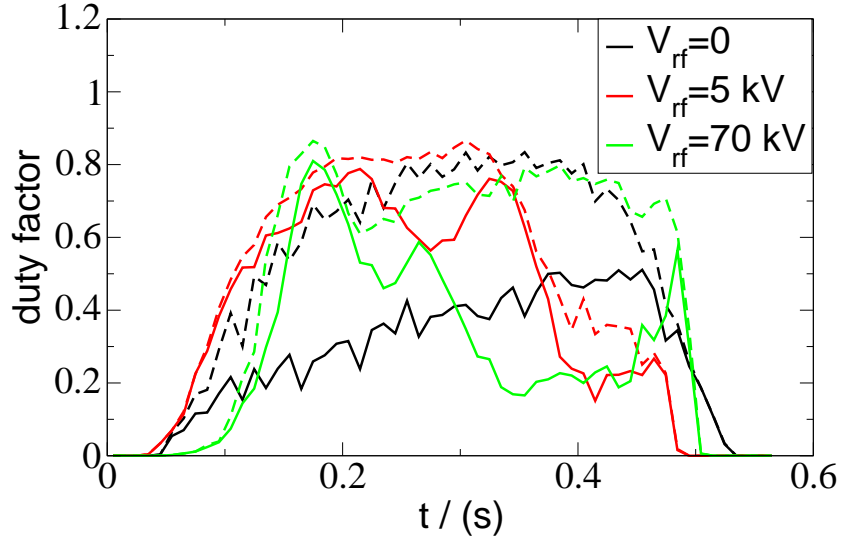}
}
\caption{Duty factors which correspond to the spills in Figures 
\ref{fig_sim_spill_ar18_500mev_vrf70kv_h100} and 
\ref{fig_sim_spill_ar18_500mev_vrf5kv_h100} compared to that obtained 
with with $V_{\rm rf}=0$, i.e. without bunching. 
The dashed curves represent the Poisson duty factors. 
}
\label{fig_sim_duty_factor_ar18_500mev_vrf70and5kv_h100}
\end{figure}

However, goal of the planned installation of the high harmonic rf cavity is 
the reduction of spill micro structures. 
Figure \ref{fig_sim_duty_factor_ar18_500mev_vrf70and5kv_h100} shows the 
time dependent duty factors obtained in simulations for rf fields with voltages 
$V_{\rm rf}=5 \ {\rm kV}$ and $V_{\rm rf}=70 \ {\rm kV}$ and the harmomic number $h=100$ 
compared with that for extraction of unbunched beam. 
The duty factors found for both rf voltages are at the beginning of the 
extraction time interval higher than that 
obtained than that obtained for unbunched beam. 
The duty factor reduction towards the end of the spill found for $V_{\rm rf}=5 \ {\rm kV}$ 
is, basically, caused by the reduction of the extraction rate which is 
denoted by the simultaneous reduction of the Poisson duty factor and which can be seen 
in Figure \ref{fig_sim_spill_ar18_500mev_vrf5kv_h100}. 
In contrast to that, the Poisson duty factor found for $V_{\rm rf}=70 \ {\rm kV}$ 
remains large to the end of the spill and no reduction of the spill 
at the half extraction time is visible in Figure \ref{fig_sim_spill_ar18_500mev_vrf70kv_h100}. 
Hence, that duty factor reduction arises from an increase of the 
micro spill structure level. 
These results indicate that an improvement of the spill quality on micro structure 
level for the rf frequency increased by an order of magnitude is possible, where 
the spill micro structure mitigation is more efficient when applying a lower 
rf voltage. 

\section{Summary}

Bunched beam extraction is a known technique for mitigating the formation of 
spill micro structures which arise from ripples in the quadrupole magnets 
of ring accelerators. 
The quadrupole ripples cause tune ripples and, thereby, a modulation of the 
size of the stable phase space area. 
The resulting structures imprinted on the flow of particles across the 
edge of the stable phase space area are reduced by synchrotron motion because it 
decreases the duration of stay of particles near the oscillating separatrices. 

We investigated the validity of spill micro structure improvement by bunched beam 
extraction for larger synchrotron tunes. 
That was primarily motivated by the plan to install an rf cavity with an rf 
frequency of $f_{\rm rf}=80 \ {\rm MHz}$ in the GSI heavy ion synchrotron SIS18 
which is approximately 20 times the rf frequency of the present cavity. 

We found spill improvement only up to an optimum synchrotron tune 
and diminution of the spill quality for increasing synchrotron tune beyond the optimum. 
The latter arises from the recapture of particles due to the synchrotron motion. 
That recapture reduces the transit time of the particles, where 
average and spread of the transit times are more reduced for a higher synchrotron tune. 
Therefore, the spill quality defined by micro structures becomes worse at higher cavity 
voltages. 

In addition, {\bf fast} synchrotron motion is found in simulations and spill measurement 
to cause the formation of macroscopic spill structures of duration $\sim 100 \ {\rm ms}$. 
Crossing synchro-betatron resonances during the tune sweep could be identified  as the 
source. 
The structures are better visible for a higher synchrotron tune because the distance 
between the resonances is proportional to the synchrotron tune. 
Consequently, the structures are closer to each other and overlap for lower 
synchrotron tune such that they are smeared out. 
Hence, inhibiting the formation of macroscopic structures low can be achieved by 
setting the rf voltage as low as possible. 
Further, it was shown that the formation of the macroscopic structures can be reduced 
by increasing the momentum width or the transverse emittance of the direction used 
for the extraction, where the latter has a {\bf strong}er impact. 

Generally, the formation of macroscopic spill structures as well as the  diminution 
of spill quality mark, in principle, a limitation to the possibility to mitigate 
spill micro structures by increasing the synchrotron tune in order to reduce the 
duration of stay of particles near the separatrix. 

\end{document}